\pdfoutput=1

\documentclass[prl,twocolumn,showpacs,amsmath,amssymb,floatfix,superscriptaddress]{revtex4}
\usepackage{graphicx}
\usepackage{dcolumn}
\usepackage{bm}
\usepackage{amsmath}
\usepackage{amssymb,xcolor,soul}
\newcommand*{\Tr}
{\mathrm{Tr} }
\usepackage[normalem]{ulem}
\begin{document}
	
	\title{Microscopic Origin of Spin-Orbit Torque in Ferromagnetic Heterostructures: A First Principles Approach}
	
	\author{Farzad Mahfouzi}
	\email{farzad.mahfouzi@gmail.com}
	\affiliation{Department of Physics and Astronomy, California State University, Northridge, CA, USA}
	\author{Rahul Mishra}
	\affiliation{Department of Electrical and Computer Engineering, National University of Singapore, 117576, Singapore}
	\author{Po-Hao Chang}
	\affiliation{Physics Department, University of Texas at El Paso, El Paso, Texas 79968, USA}	
	\author{Hyunsoo Yang}
	\affiliation{Department of Electrical and Computer Engineering, National University of Singapore, 117576, Singapore}
	\author{Nicholas Kioussis}
	\email{nick.kioussis@csun.edu }
	\affiliation{Department of Physics and Astronomy, California State University, Northridge, CA, USA}
	
	\begin{abstract}
We present an {\it ab initio}-based theoretical framework which elucidates the origin of the spin-orbit torque (SOT) in  Normal-Metal(NM)/Ferromagnet(FM) heterostructures. The SOT is decomposed into two contributions, namely, {\it spin-Hall} and the {\it spin-orbital} components. We find that 
{\it (i)} the Field-Like (FL) SOT is dominated by the spin-orbital component and {\it (ii)} both components contribute to the damping-like torque with comparable magnitude in the limit of thick Pt film. The contribution of the spin-orbital component to the DL-SOT is present only for NMs with strong SOC coupling strength. We demonstrate that the FL-SOT can be expressed in terms of the non-equilibrium spin-resolved  orbital moment accumulation.
The calculations reveal that the experimentally reported oxygen-induced sign-reversal of the FL-SOT in Pt/Co bilayers is due to the significant reduction of the majority-spin orbital moment accumulation on the interfacial NM atoms.

	\end{abstract}
	
	\date{\today}
	\maketitle
	
	Spin-Orbit torque (SOT) has recently attracted a lot of attention as a method to switch nano-scale magnetic bits, due to its promising features in terms of high efficiency and scalability.\cite{Manchon2008,Miron2011,Liu2012,Liu2012_1,Cubukcu2014,CZhang2015,Miron2010,Garello2013,Haney2013,Lee2015,Freimuth2014,MahfouziPRB2018,Filipe2017,WimmerPRB2016,Kirill2019} 	
	SOT is a relativistic phenomena that has its origin in the atomic spin-orbit coupling (SOC) in systems with broken inversion symmetry, and is often separated into damping (Slonczewski)-like (DL), $\tau_{DL} \vec{m}\times(\vec{m}\times\vec{y})$, and field-like (FL), $\tau_{FL} \vec{m}\times\vec{y}$, components, where $\vec{m}$ is the unit vector along the magnetization direction and $\vec{y}$ is the in-plane unit vector normal to the external electric field. 
	The origin of the SOT is conventionally attributed to extrinsic  and intrinsic effects. First principles electronic structure calculations of SOT using the coherent potential approximation\cite{WimmerPRB2016}  suggest negligible contribution from extrinsic mechanisms({\it e.g.} skew-scattering and side-jump mechanisms). The intrinsic contribution is conventionally attributed to  spin-Hall\cite{SinovaRMP2015} (SH) and Rashba-Edlestein\cite{Edelstein1990} (RE) effects.
	
	The RE effect is understood in terms of the spin-momentum locking of the electronic Bloch states, where a non-zero expectation value of the momentum in the presence of a current flow leads to a finite spin accumulation in the FM. The resulting spin accumulation couples to the local magnetic moments through the exchange splitting, leading to a current-induced FL torque on the magnetization direction of the FM. The RE mechanism of the FL-SOT is often phenomenologically modeled by a 2D plane with SOC introduced through the Rashba term in the Hamiltonian of the form, $m_e\alpha_R(\vec{v}_{{k}}\times\vec{\hat{\sigma}})\cdot\vec{e}_z$, where $\vec{v}_k$ is the electronic group velocity, $\vec{e}_z$ is a unit vector normal to the Rashba plane, $m_e$ is electron's mass and $\alpha_R$ is the  Rashba coefficient.
	This model yields an effective current-induced magnetic field, $\vec{B}^{eff}_{SOT}=2m_e\alpha_{R}\vec{I}_S\times\vec{e}_z/M_s$, where $\vec{I}_S$ is the in-plane spin current polarized along the magnetization direction and $M_s$ is the magnetic moment of the FM. \cite{WangPRB2014,Manchon2009,Kalitsov2017,Chotorlishvili2019} For the actual NM/FM bilayer, the same model is usually employed by introducing the Rashba SOC localized at the interfacial region. 
	In addition to the FL-SOT, the RE effect can also give rise to DL-SOT.\cite{SinovaRMP2015,ManchonRMP2018,AminPRB2016,WangPRB2014,
	Li2015,QaiumzadehPRB2015} 

 Unlike the RE effect which is considered to be a {\it local} (interfacial) phenomena, the SH mechanism of the SOT, on the other hand, can be viewed as a {\it non-local} effect. In the SH mechanism, the NM and FM films are assumed as two independent entities where the spin current is generated in the bulk NM and subsequently absorbed by the FM. The efficiency of charge- to spin-current conversion is measured 
by the spin-Hall angle, $\Theta_{SH}$, which is given by, $\vec{I}_{\vec{S}}=\frac{\hbar}{2e}\Theta_{SH}\vec{I}_C\times\vec{e}_{\vec{S}}$, 
where $\vec{e}_{\vec{S}}$ is the spin polarization unit vector and $\vec{I}_C$ is the charge current. The spin current in turn interacts with the magnetic moment of the FM, resulting in DL and FL-SOTs.\cite{Haney2013,ManchonJoP2016,WangPRB2014} 
The SH mechanism of SOT is often treated within the spin drift-diffusion model\cite{SinovaRMP2015,Dyakonov1971}, where RE effect is usually introduced to model the spin memory loss at the NM/FM interface \cite{Rojas2014,DoluiPRB2017,Bass2007, KirillPRL2016, MahfouziPRB2012}. Furthermore, additional mechanisms such as, the interface-generated spin-Hall current in the FM layer \cite{AminPRL2018} and orbital-Hall effect \cite{Dongwook2018} have also been suggested to contribute to the DL-SOT.
	
	Although both SH and RE mechanisms offer simple and qualitative pictures for the origin of the SOT, the SH effect alone is typically insufficient in providing a quantitative assessment of the experimental measurements\cite{ManchonRMP2018,Manchon2008,SinovaRMP2015}; and the RE effect has proven useful mostly in phenomenological investigations of the SOT. Thus, there is an urgent need for a microscopic framework to explain the origin of the SOT in terms of the electronic structure properties of the materials involved and the chemistry of the interface.\cite{HellmanRMP2017} 
	
	In this letter, we develop an ab-initio based formalism, where the SOT is decomposed into {\it spin-Hall} (non-local) and {\it spin-orbital} (local) components. We show that the FL-SOT is dominated by the {\it spin-orbital} component originating from the interfacial Pt layer, while both components contribute on equal footing to the DL-SOT. We demonstrate that the spin-orbital component of the FL-SOT is related to the non-equilibrium spin-resolved orbital moment accumulation on the normal metal. The physical meaning of the {\it spin-orbital} vs {\it spin-Hall} decomposition is systematically studied by the Pt thickness dependence, the layer-resolved contribution and connection to the spin current passing through the NM/FM interface. 
	
 Transport properties, including the SOT, can be determined within the Green function formalism 
  either in real space \cite{Kirill2019,Mahfouzi_PRB2016} using the Landauer-like approach or equivalently in momentum space using the Kubo-like approach, \cite{Freimuth2014,MahfouziPRB2018,WimmerPRB2016}. In these approaches the current-induced spin accumulation induces SOT,	
	\begin{align}\label{eq:Eq1}
	\vec{\tau}_{sot}=2\vec{m}\times\langle\hat{\Delta}_{ex}\vec{\hat{\sigma}}\rangle_{neq}/M_s,
	\end{align}	
	 on the magnetization direction $\vec{m}$, through the magnetic exchange splitting of the conduction electrons, $\hat{\Delta}_{ex}=(\hat{H}_{\uparrow\uparrow}-\hat{H}_{\downarrow\downarrow})/2$. 
	 Here, $\hat{H}_{\sigma\sigma}$ is the electronic Hamiltonian  for spin $\sigma$, $M_s$ is the magnetic moment per unit cell and $\langle...\rangle_{neq}$ denotes the nonequilibrium expectation value. Although, this approach has proven advantageous in producing reasonable results in comparison with experimental measurements\cite{MahfouziPRB2018,Kirill2019,Mishra2019}, it does not offer the means to directly analyze the microscopic origin of the SOT in terms of the electronic structure of the heterostructure. 
	 
	As an alternative to the spin density calculation approach, here we use Hamilton's equations of motion for the canonical variables, 
	$\phi,\theta$ (see Fig. S1 in Ref.~\cite{Supp_Matt}) to determine the Fermi surface contribution to the non-equilibrium canonical forces,
	\begin{align}\label{eq:EqoMs2}
	F_q&=\frac{2e E^x_{ext}}{M_sN_k\pi}\sum_{\vec{k}}Im\Tr\Big(\hat{\eta}\frac{\partial \hat{G}_{\vec{k}}}{\partial q}\hat{v}_{k_x}\hat{G}^{\dagger}_{\vec{k}}\Big),\ \ \ (q=\phi,\theta),
	\end{align}
where, $E^x_{ext}$ is the external electric field along $x$, $N_k$ is the number of k-points in the unit cell, 
$\hat{G}_{\vec{k}}$ is the Green's function, 
$\hat{v}_{k_{x}}$ is the electronic group velocity
and $\hat{\eta}=\eta\hat{1}$ is the energy broadening parameter. Rotating the 
reference frame so that the magnetization's orientation is along the $z$-axis, the partial derivative of the Greens function can be written as 	
	\begin{align}\label{eq:EqoM3}
	\hat{U}_{\vec{m}}^{\dagger}\frac{\partial \hat{G}_{\vec{k}}}{\partial q}\hat{U}_{\vec{m}}&=	i[\hat{\mathcal{G}}_{\vec{k}},\hat{O}_q]+\hat{\mathcal{G}}_{\vec{k}}\frac{\partial \hat{H}^{rot}_{SOC}}{\partial q}\hat{\mathcal{G}}_{\vec{k}}, 
	\end{align}
	where, the spin rotation operator, $\hat{U}_{\vec{m}}=e^{i(\vec{n}\cdot\hat{\vec{\sigma}})\theta/2}$, was used on the Hamiltonian 
	to align the magnetization along the fixed $z$-direction.
	Here,  $\vec{n}=\cos(\phi)\vec{e}_y-\sin(\phi)\vec{e}_x$,  $\vec{e}_{x,y,z}$ are unit vectors along the Cartesian coordinates,
	$ \hat{\mathcal{G}}_{\vec{k}}$ and 
	$\hat{H}^{rot}_{SOC}=\hat{U}_{\vec{m}}^{\dagger}\hat{H}_{SOC}\hat{U}_{\vec{m}}$  are the Green's function and SOC Hamiltonian in the rotated frame, respectively, 
	$\hat{O}_q=i\hat{U}_{\vec{m}}^{\dagger}\partial \hat{U}_{\vec{m}}/\partial q$, 
		and 
	$\hat{H}_{SOC} = \chi\hat{\xi}\hat{\vec{L}}\cdot\hat{\vec{\sigma}}$, where $\hat{\vec{L}}$ is the angular momentum operator, $\hat{\vec{\sigma}}$'s are the Pauli matrices, $\chi$ is the SOC scaling factor, and $\hat{\xi}$ is the SOC matrix.
	
	The first term in Eq.~(\ref{eq:EqoM3}) contributes only to the non-equilibrium observables and often describes the non-local spin-current pumping/absorption effects\cite{MahfouziPRB2012,Supp_Matt,mahfouziPRB2017_GD}. On the other hand, the second term generally leads to the contribution from modification of the band-structure in response to the change of $q$, similar to Kamberský's breathing Fermi surface mechanism of Gilbert damping\cite{mahfouziPRB2017_GD} and magnetocrystalline anisotropy using the ``torque'' method\cite{TrqMethod}.
	Using Eq.~(\ref{eq:EqoM3}) the non-equilibrium canonical force, $F_{q}$, can be decomposed into the following two components, which we refer to as the {\it spin-orbital} and {\it spin-Hall} contributions, respectively,\cite{Supp_Matt}
   
   \begin{subequations}
   	\begin{align}
   	&F_{q}^{(so)}=\frac{2e E^x_{ext}}{M_sN_k\pi}\sum_{\vec{k}}Im\Tr\Big[\frac{\partial\hat{H}^{rot}_{SOC}}{\partial q}\hat{\mathcal{G}}_{\vec{k}}\hat{v}_{k_x}Im(\hat{\mathcal{G}}_{\vec{k}})\Big], 
   	\label{eq:EqoMs3a_0}\\
   	&F_q^{(sh)}=\frac{2eE^x_{ext}}{M_sN_k\pi} \sum_{\vec{k}}Re\Tr(\hat{\eta}[\hat{O}_q,\hat{\mathcal{G}}_{\vec{k}}]\hat{v}_{k_x}\hat{\mathcal{G}}^{\dagger}_{\vec{k}}). \label{eq:EqoMs3b_0}
   	\end{align}	
   \end{subequations}

The canonical forces $F_{q=\theta,\phi}$ are related to the torque through $F_{\phi}=\vec{\tau}_{sot}\cdot\vec{e}_z$ and $F_{\theta}=\vec{\tau}_{sot}\cdot\vec{n}$.
To the lowest order in the angular dependence of the SOT, we expect,  $\vec{\tau}_{sot}=\vec{m}\times(\vec{B}_{FL}+\vec{m}\times\vec{B}_{DL})$. 
The magnitude of the SOTs, $\vec{B}_{FL,DL}$, can then be calculated by fitting the angular dependence of the canonical forces to the expressions,
	 \begin{subequations}\label{eq:EqoMs2a_0}
	 	\begin{align}
	 	F_{\phi}^{\alpha}/E^x_{ext}& = \vec{m}\times(\vec{B}^{\alpha}_{FL}+\vec{m}\times\vec{B}^{\alpha}_{DL})\cdot\vec{e}_z,\\
	 	F_{\theta}^{\alpha}/E^x_{ext}&=
	 	\vec{m}\times(\vec{B}^{\alpha}_{FL}+\vec{m}\times\vec{B}^{\alpha}_{DL})\cdot\vec{n},
	 	\end{align}	
	 \end{subequations}  
where $\alpha$ refers to the spin-orbital or spin-Hall contributions to the SOT.


We have also derived analytical expressions for the spin-orbital/spin-Hall components of the FL- and DL-SOTs\cite{Supp_Matt}. The FL-SOT is of the form
\begin{align}\label{eq:EqoMs4_0}
\frac{M_s}{2e}&\vec{B}^{(so)}_{FL}	=\langle\hat{\xi}\vec{\hat{L}}\hat{{\sigma}}_z\rangle_0-\langle\hat{\xi}{\hat{L}}_z\vec{\hat{\sigma}}\rangle_0-\langle\hat{\xi}\vec{\hat{L}}\cdot\hat{\vec{\sigma}}\rangle_0\vec{e}_z . 
\end{align}
where, 
\begin{align}
\langle... \rangle_0&=\frac{1}{N_k\pi}\sum_{\vec{k}}\Tr(Im(\hat{\mathcal{G}}_{\vec{k}})\hat{v}_{k_x}Im(\hat{\mathcal{G}}_{\vec{k}})...),
\end{align}	
Even though,  $\vec{B}^{(so)}_{FL}$, has three contributions we find that in the case of NM/FM bilayer devices the first term, which we refer to as the Rashba-Edlestein Effect (REE) FL-SOT, 
 \begin{align}\label{eq:Eq6}
&\frac{M_s}{2e}\vec{B}^{REE}_{FL}  \equiv \langle\hat{\xi}\vec{\hat{L}}\hat{{\sigma}}_z\rangle_0  = \langle\hat{\xi}\vec{\hat{L}}\rangle_{0,\uparrow}-\langle\hat{\xi}\vec{\hat{L}}\rangle_{0,\downarrow}, 
\end{align}
is dominant, while the third term is present only in systems with broken in-plane mirror symmetry.\cite{Yu2014,MacNeill2017} Eq.~\eqref{eq:Eq6} is one of the {\it central} results of this letter which demonstrates that the FL-SOT can be expressed in terms of {\it the non-equilibrium spin-resolved  orbital moment accumulation}. 
Similarly, the spin-Hall component of the DL-SOT is given by, 
\begin{align}\label{eq:DL_SHE_0}
\frac{M_s}{e}\vec{B}^{(sh)}_{DL}&=-\frac{1}{\pi N_k} \sum_{\vec{k}}Re\Tr(\hat{\eta}[\vec{\hat{\sigma}},\hat{\mathcal{G}}_{\vec{k}}]\hat{v}_{k_x}\hat{\mathcal{G}}^{\dagger}_{\vec{k}}).
\end{align}	
These expressions allow to elucidate the microscopic origins of the SOT. More specifically,  Eq.~\eqref{eq:EqoMs4_0} will be employed to understand the interfacial Co oxidation effect on the  FL-SOT in the Pt/Co bilayer. 
Eq.~\eqref{eq:EqoMs4_0} can also be used to estimate the effective Rashba SOC strength of the bilayer which is given by $m_e\alpha^{eff}_R=\langle\hat{\xi}\hat{L}_y\rangle_0/\langle\hat{v}_{k_x}\rangle_0$.\cite{Supp_Matt}

\begin{figure} [tbp]
		{\includegraphics[angle=0,trim={0cm 6.9cm 1.0cm 2.0cm},clip,width=0.25\textwidth]{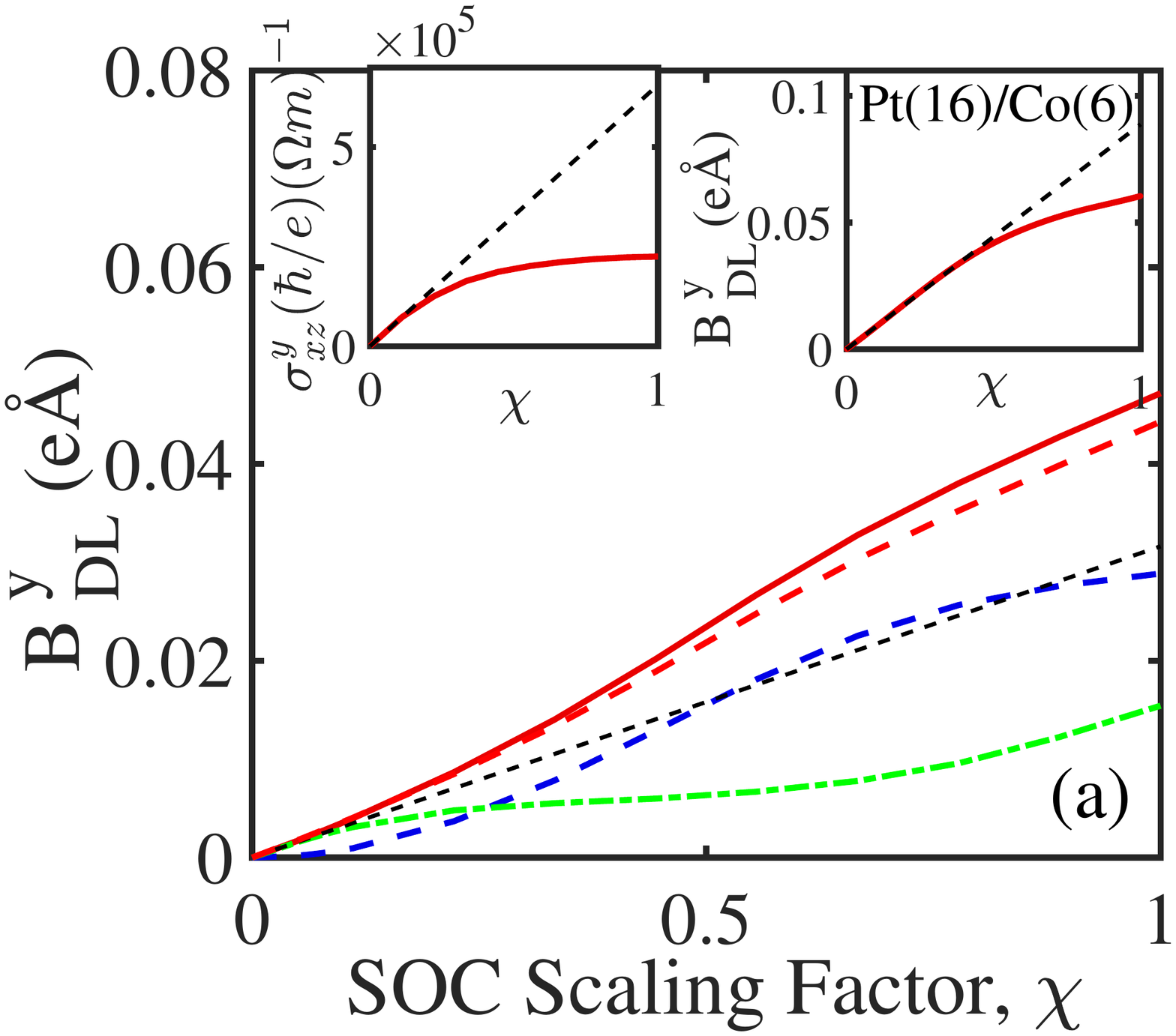}}%
		{\includegraphics[angle=0,trim={0cm 6.9cm 1.0cm 2.0cm},clip,width=0.25\textwidth]{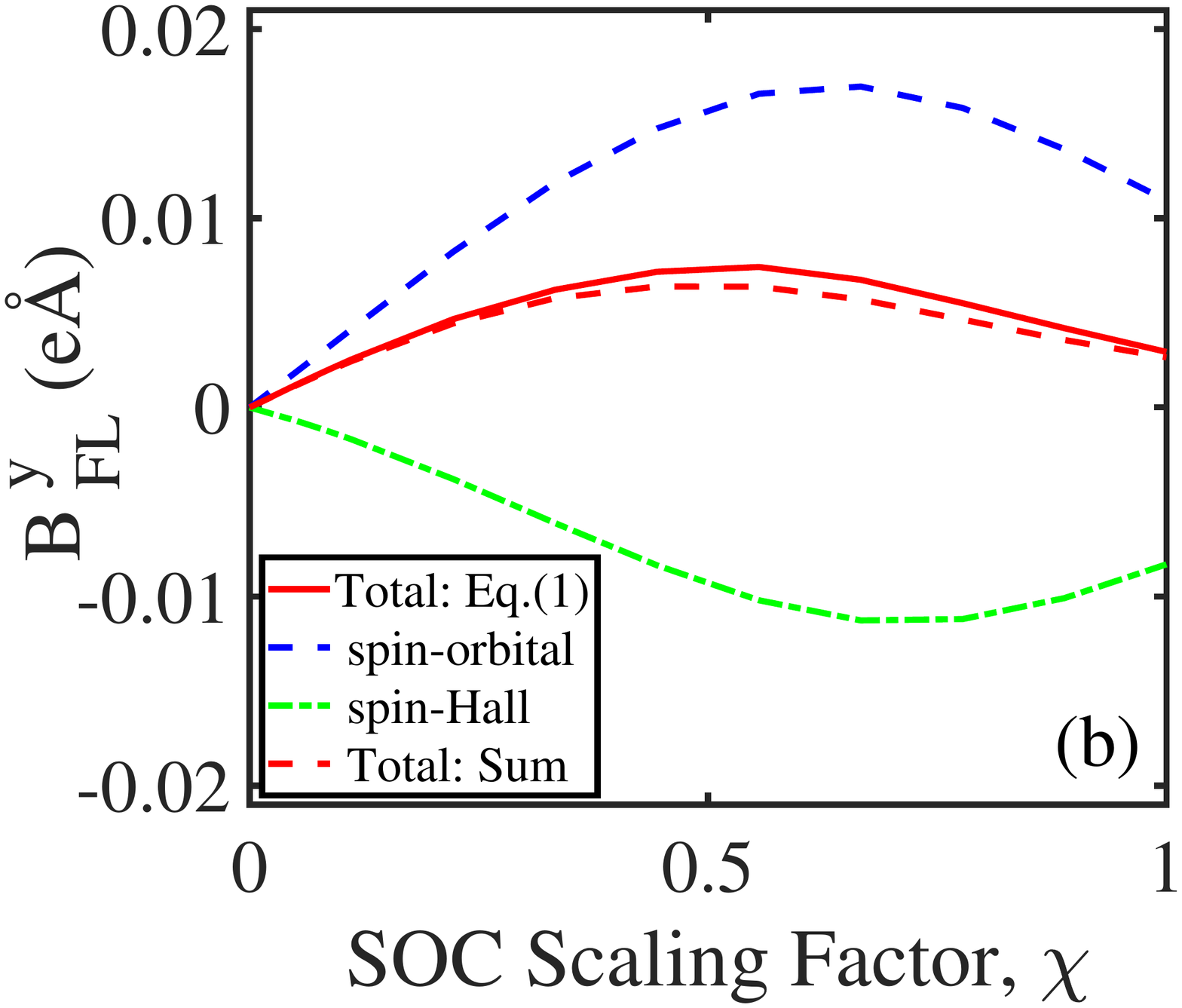}}%
		\caption{(Color online) 
	Total (solid red curves) (a) DL and (b) FL  components of the SOT, calculated from Eq.~\eqref{eq:Eq1},  versus the SOC scaling factor $\chi$, for the (001) Pt(6 ML)/Co(6 ML) bilayer with 
 energy broadening $\eta=0.1 eV$.
We also show the spin-orbital (dashed blue) and spin-Hall (dash-dotted green), contributions to the SOT calculated from Eq.~\eqref{eq:EqoMs2a_0}, their sum (dashed red) and its linear dependence (black line in (a)). 		
Left inset: Spin-Hall conductivity of bulk Pt versus $\chi$. Right inset: DL-SOT vs $\chi$ for thicker Pt film in the Pt(16 ML)/Co(6 ML) bilayer.}
		\label{fig:fig2}
	\end{figure}

In a NM/FM bilayer, only the $y$-component of $\vec{B}_{FL,DL}$  is nonzero. Fig.~\ref{fig:fig2} shows the total (solid red curves) (a) DL and (b) FL  components of the SOT, calculated from Eq.~\eqref{eq:Eq1},  versus the SOC scaling factor 
	$\chi$, for the (001) Pt(6)/Co(6) bilayer [see Fig.~\ref{fig:fig4}b]. Here, the integers in the parenthesis denote the number of monolayers (MLs). We also show both the spin-orbital (blue) and spin-Hall (green) contributions to the SOT, calculated from Eq.~\eqref{eq:EqoMs2a_0}, and their sum (red dashed curve). Within the accuracy of the numerical calculations and errors in the fitting of the angular dependence, overall, there is good agreement of the results calculated from 
	two different aforementioned approaches, namely Eq.~\eqref{eq:Eq1} (red solid curve) and the sum of  Eqs. 3(a) and (b). 
In the limit of $\chi\rightarrow$0, the spin-Hall contribution to the DL-SOT in Fig.~\ref{fig:fig2}(a) varies linearly with $\chi$ (as expected from the conventional spin-Hall effect), while the spin-orbital contribution exhibits a quadratic behavior with SOC (See Ref.~\cite{Supp_Matt} for the analytical derivation/discussion).
With increasing SOC the spin-Hall contribution is reduced from its linear dependence (dashed black line) similar to the behavior of the spin-Hall conductivity in bulk Pt shown in the left inset of Fig.~\ref{fig:fig2}(a).
This is due to the  enhancement of the spin dephasing effects in bulk Pt and the interfacial region\cite{DoluiPRB2017,KirillPRL2016}. On the other hand, the dominant contribution from the spin-orbital component to the DL-SOT for thin Pt film (6 MLs) leads to enhancement of the total SOT relative to its linear behavior.  
We also show for comparison in the right inset the variation of the total DL-SOT for thicker (16 MLs) Pt film. The slight suppression of the total DL-SOT from its linear behavior for thicker Pt films, indicates the crucial role of the spin-orbital component for unltra-thin Pt films. 	
	
	The spin-Hall (green curve) and spin-orbital (blue) contributions to the FL-SOT in Fig.~\ref{fig:fig2}(b) exhibit a linear behavior versus SOC in the limit  of $\chi\rightarrow$0. Interestingly, 
	the two contributions have opposite sign.  
	The dependence of the spin-orbital and spin-Hall contributions to the FL-SOT on 
	the energy broadening parameter, $\eta$, (inversely proportional to the relaxation time)\cite{Supp_Matt}, shows that in the ballistic limit ($\eta\rightarrow$0) $\vec{B}_{FL}^{(so)}\propto1/\eta$ while $\vec{B}_{FL}^{(sh)}\propto\eta$, suggesting that 
	the latter can be ignored in relatively clean samples. 
	

	\begin{figure}		
		{\includegraphics[angle=0,trim={0.3cm 7.0cm 0cm 3cm},clip,width=0.25\textwidth]{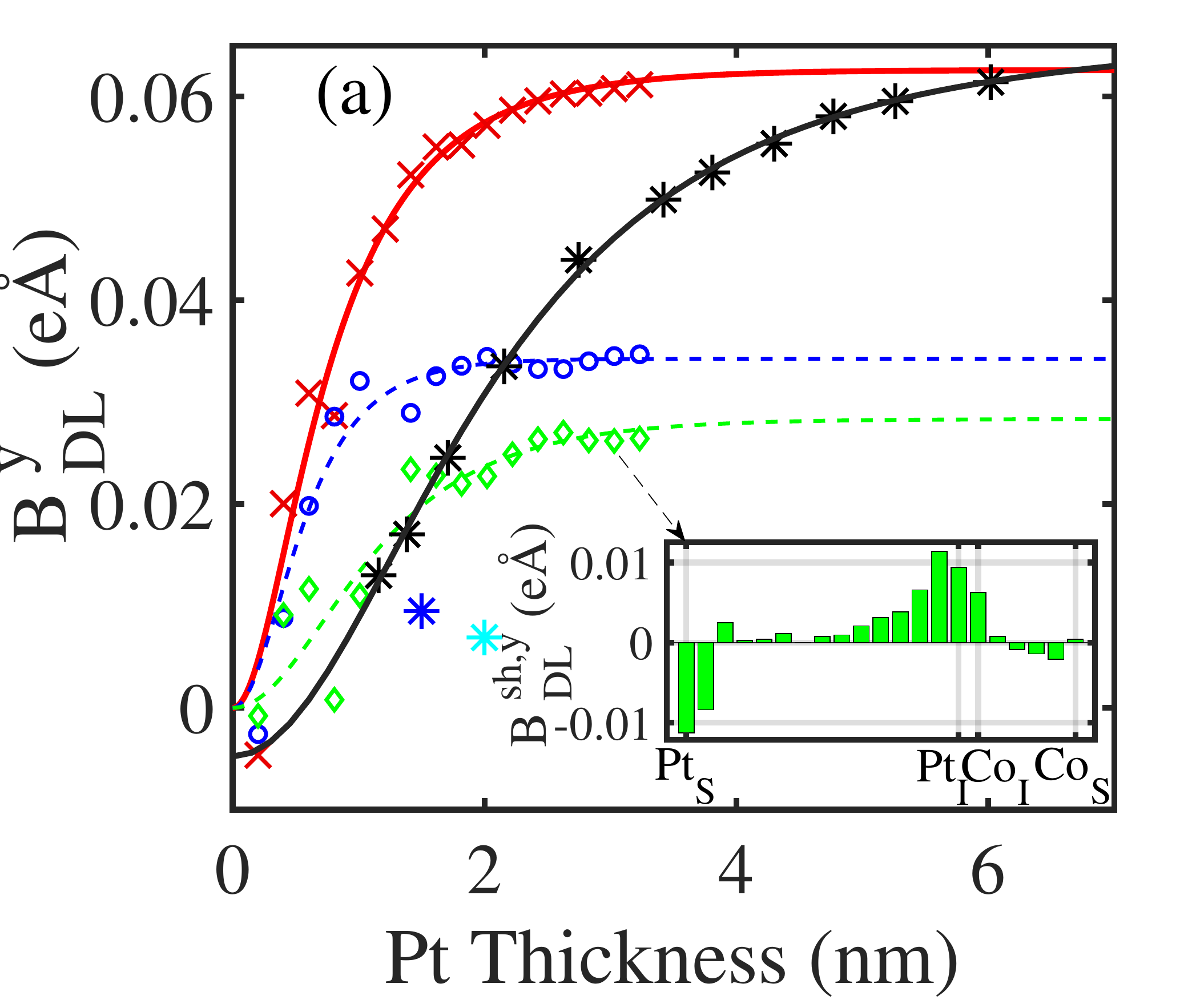}}%
		{\includegraphics[angle=0,trim={0.3cm 7.0cm 0cm 3cm},clip,width=0.25\textwidth]{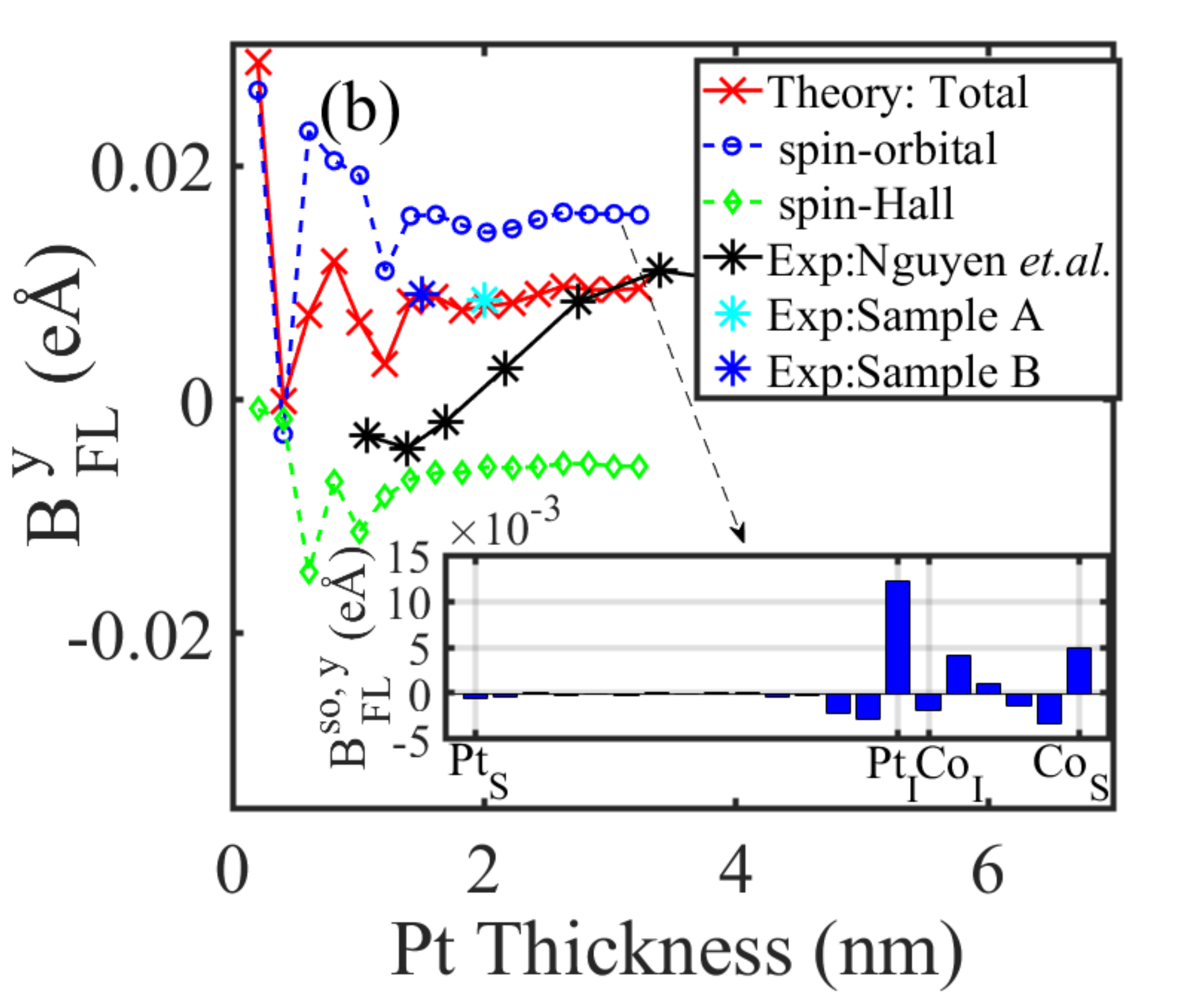}}%
	\caption{(Color online) 
Total (a) DL-SOT and (b) FL-SOT (red crosses) calculated from Eq.~\eqref{eq:Eq1} versus Pt thickness for $\eta$=0.1 eV and compared with the experimental results (multiplied by $\mu_B$) shown as black (Ref.~\cite{Nguyen2016}) and dark and light blue (Ref.~\cite{Mishra2019}) stars. Blue circles and green diamonds denote the spin-orbital and spin-Hall contributions to the SOT calculated from Eq.~\eqref{eq:EqoMs2a_0}, respectively.  	
The black solid and dashed curves in Fig.~\ref{fig:fig3}(a) are fits of the DL-SOT to the spin-diffusion model\cite{SinovaRMP2015,Dyakonov1971} while the red solid curve is the sum of the two dashed lines. Insets in (a) and (b): Layer-resolved contribution to the DL- and FL-SOT for 15 MLs Pt thickness.}
	\label{fig:fig3}
	\end{figure}

A convenient approach to characterize the spin-Hall and spin-orbital components in terms of their bulk and interface contributions is to investigate their dependence on the HM film thickness, where the interfacial component is expected to have a shorter characteristic length (i.e. spin diffusion length). 
In Fig.~\ref{fig:fig3}(a)  we display the total (red crossed symbols) DL-SOT, calculated from the nonequilibrium spin accumulation method\cite{MahfouziPRB2018}, Eq.~\eqref{eq:Eq1}, 
versus Pt thickness, along with the experimental results (black~\cite{Nguyen2016} and blue\cite{Mishra2019} stars)
(multiplied by $\mu_B$).
We also show both the spin-orbital (blue circles) and spin-Hall (green diamonds) contributions to the SOT calculated from Eq.~\eqref{eq:EqoMs2a_0}. 	
The blue and green dashed curves in Fig.~\ref{fig:fig3}(a) denote the fits of the {\it ab initio} results to the spin-diffusion model,  
$\propto [1-sech(d_{Pt}/\lambda)]$,\cite{SinovaRMP2015,Dyakonov1971}, while, the solid red curve is the sum of the two dashed curves. Here $d_{Pt}$ is the Pt thickness and $\lambda$ is the effective 
spin diffusion length in Pt. We find that $\lambda^{(so)}=0.4\ nm$ and $\lambda^{(sh)}=0.7\ nm$, both of which are in agreement with the reported experimental values in the range between 0.5 and 10 nm.\cite{Rojas2014,Boone2013} 
It is also worth pointing out that, Fig.~\ref{fig:fig3}(a) shows a sign reversal of both the spin-orbital and spin-Hall DL-SOT components for 1 ML Pt, indicating the crucial role of the local electronic structure on the strength and sign of the SOT. A similar sign reversal for thin layer of Hafnium \cite{Ramaswamy2016} and Tantalum \cite{Kim2013} have already been reported experimentally. The layer-resolved contribution of ${B}^{(sh),y}_{DL}$, displayed in the 
inset of Fig.~\ref{fig:fig3}(a) for 15 MLs of Pt, demonstrates the dominant bulk origin of the DL-SOT with a negative contribution from the surface Pt layer. The relatively significant contribution from the surface Pt in the case of ultrathin Pt films suggests the sensitivity of the DL-SOT on the substrate material. The layer-resolved results were calculated from the diagonal matrix elements inside the trace in Eq. (S10) in Ref.~\cite{Supp_Matt}, with spin-orbit coupling included in the Green's function.
	

 Fig.~\ref{fig:fig3}(b) shows the total (red crosses) FL-SOT, calculated from Eq.~\eqref{eq:Eq1},
   	versus Pt thickness along with the experimental results (black stars)
   	(multiplied by $\mu_B$).~\cite{Nguyen2016}
   	We also show both the spin-orbital (blue circles) and spin-Hall (green diamonds) contributions to the FL-SOT, calculated from Eq.~\eqref{eq:EqoMs2a_0}, respectively. 
   	Note that the spin-orbital component of the FL-SOT yields the dominant 
   	contribution and that the spin-Hall component has opposite sign for $\eta$=0.1 eV. 
   	The calculated  total FL-SOT are in good agreement with the experiments~\cite{Nguyen2016,Mishra2019}.
   	The oscillation  
   	of the {\it ab initio} FL-SOT for ultrathin Pt film ($d_{Pt}<$2\ nm) presumably arises from 
   	the finite size effects. 
   	The layer-resolved contribution (calculated from diagonal matrix elements inside the trace in Eq.~\eqref{eq:EqoMs3a_0}) to $B^{y,so}_{FL}$ is displayed in the 
   	inset of Fig.~\ref{fig:fig3}(b) for 15 MLs of Pt, demonstrating that the interfacial Pt layer is primarily responsible for the FL-SOT on the Co film. 

\begin{figure}	
	{\includegraphics[scale=0.6,angle=0,trim={5cm 9.0cm 4.5cm 1.7cm},clip,width=0.7\textwidth]{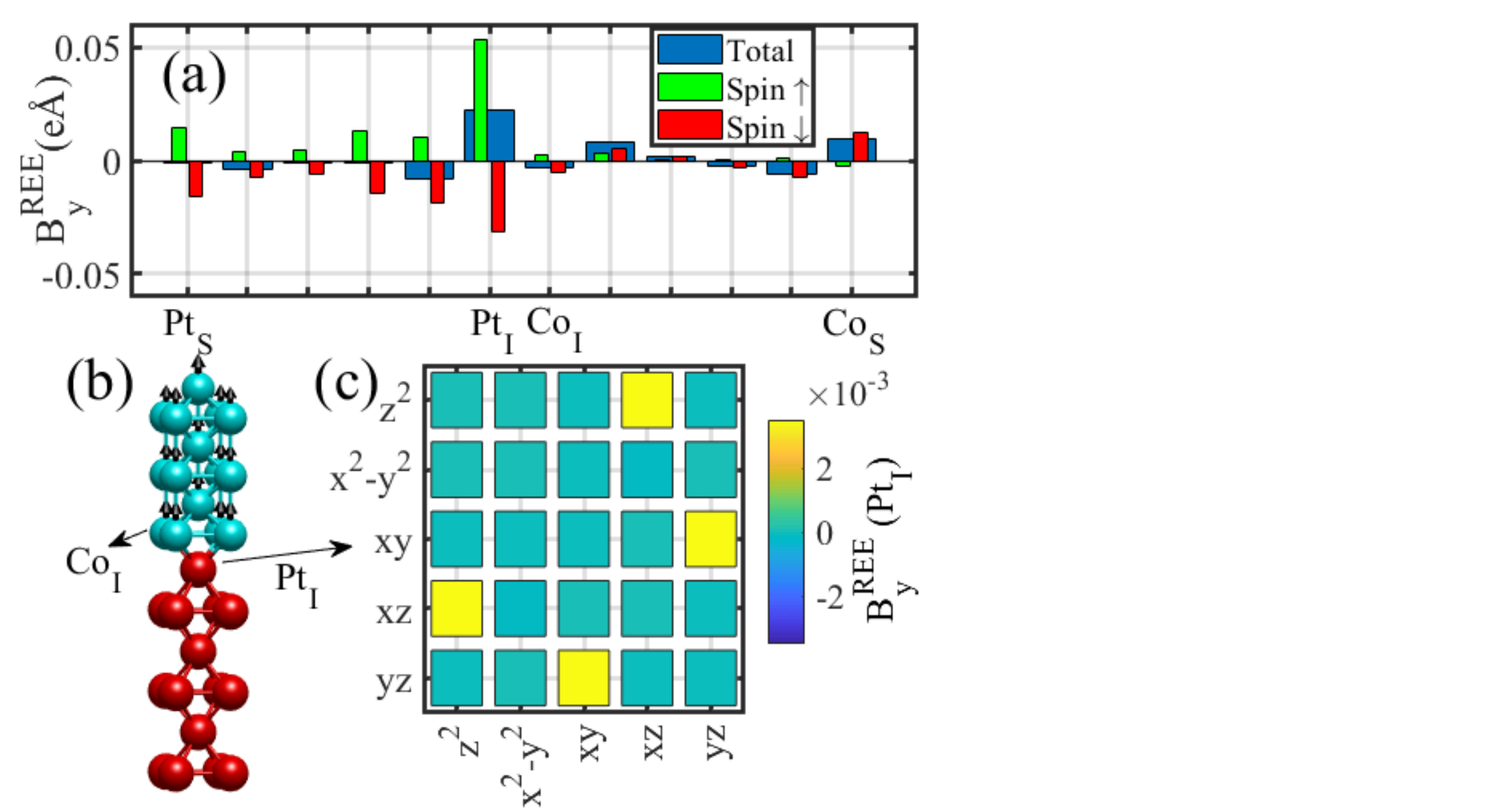}}%
	
	{\includegraphics[scale=0.6,angle=0,trim={5cm 8.9cm 4.5cm 1.9cm},clip,width=0.7\textwidth]{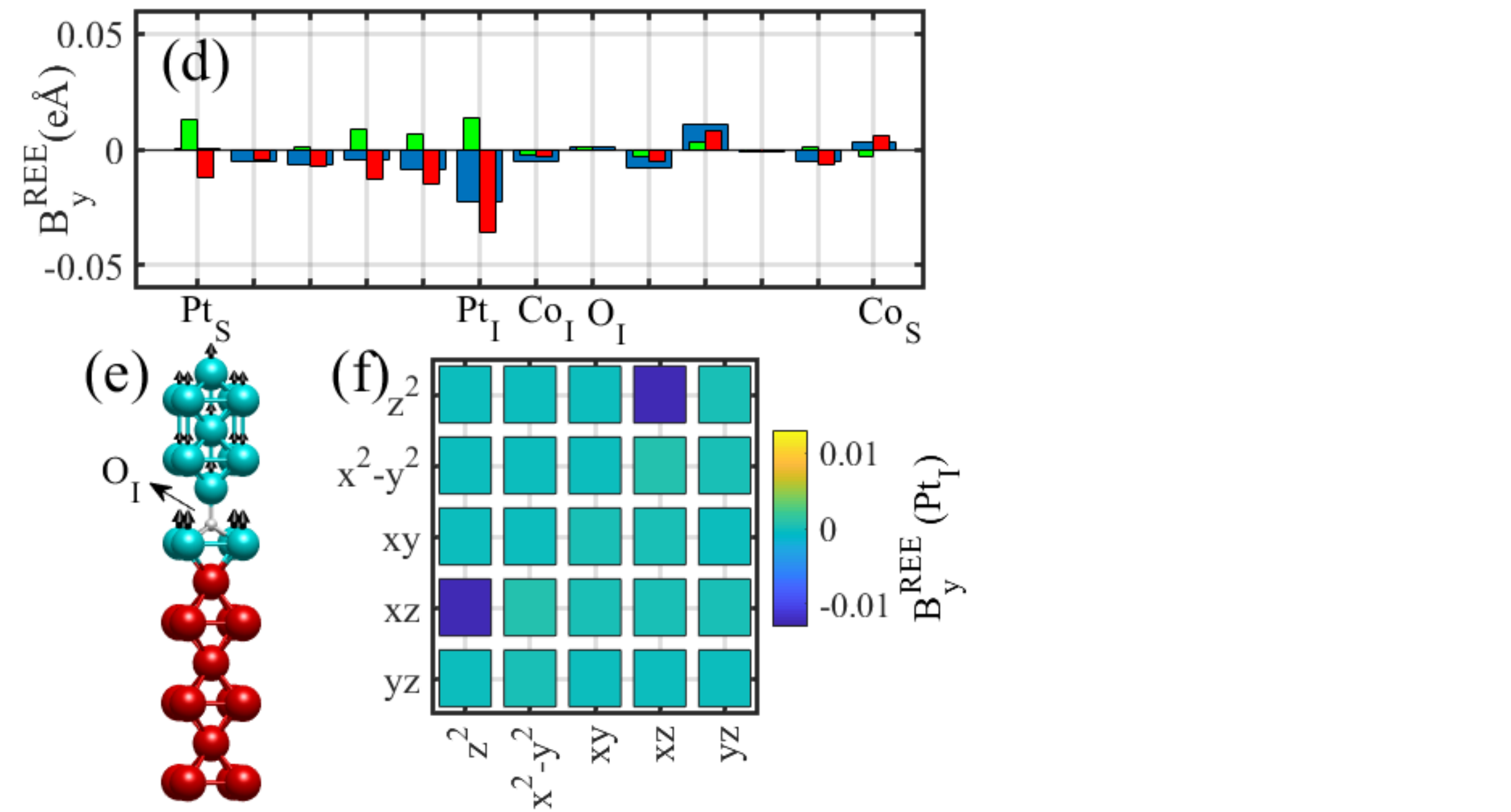}}%
	\caption{(Color online)  Spin- and layer- resolved FL-SOT in (a) the absence and (d) presence of interfacial oxygen. (c) and (f) Show the corresponding orbital-resolved FL-SOT for the interfacial Pt layer. Atomic structure of [001] Pt(6)/Co(6) bilayer without (b) and with (e) interfacial oxygen. The subscripts $I$ and $S$ denote the spin-orbital and surface layers, respectively.}
	\label{fig:fig4}
\end{figure}

{\it Effect of Interfacial Co Oxidation:} 
Recent experiments have reported\cite{Mishra2019,Qiu2015} a modulation of the direction and magnitude of the SOT in a Pt/Co/GdO$_x$ heterostructure by changing the concentration of oxygen in the Co layer using an electric field. 
Even though our complementary {\it ab initio} calculations confirmed\cite{Mishra2019}  the sign reversal of the FL-SOT  as a function of oxygen concentration, its microscopic origin has so far remained elusive. 

Here, using the theoretical framework developed above we elucidate the underlying mechanism of the sign reversal of the FL-SOT. 
We consider the (001) Pt(6 ML)/Co(6 ML) bilayer system where the oxygen atom is originally placed in the interfacial Co layer and atop of the interfacial Pt atom (Fig. \ref{fig:fig4}(e))\cite{Mishra2019}. The oxygen atom relaxes between the interface and sub-interface Co layers.  
\begin{table}[b]
	\centering
	\caption{Values of the y-component of the total FL-SOT (in meV\AA) for the (001) Co(6)/Pt(6) bilayer [Eq.~\eqref{eq:Eq1}]; the spin-orbital and spin-Hall contributions to the FL-SOT [Eq.~\eqref{eq:EqoMs2a_0}]; and the Rashba-Edlestein effect (REE) FL-SOT [Eq.~\eqref{eq:Eq6}], in the  absence and presence of interfacial oxygen, respectively.}
	\begin{ruledtabular}
		\begin{tabular}{ c | c c c r c  } 
			Co(6)/Pt(6)  &    $B_{y,FL}$   &  $B^{(sh)}_{y,FL}$   & $B^{(so)}_{y,FL}$       & $B^{REE}_{y,FL}$        \\ 
			&    Eq.~\eqref{eq:Eq1}    &       Eq.~\eqref{eq:EqoMs2a_0}         & Eq.~\eqref{eq:EqoMs2a_0}    &  Eq.~\eqref{eq:Eq6}              &                        \\
			\hline
			Absence of O & 3     & -8   & 11    &  15      \\ 
			Presence of O &   -43      & 9   & -51    & -51       \\ 
		\end{tabular}
	\end{ruledtabular}
\end{table}
Table I lists the  total FL-SOT calculated from Eq.~\eqref{eq:Eq1}, the spin-orbital and spin-Hall contributions to the FL-SOT calculated from Eq.~\eqref{eq:EqoMs2a_0}, and the REE FL-SOT calculated from Eq.~\eqref{eq:Eq6},  in the  absence and presence of interfacial oxygen, respectively, which show the oxygen-induced sign reversal. 
Figs.~\ref{fig:fig4}(a) and (d) show the layer- and spin-resolved contribution to $B_{y}^{REE}$, in the absence and presence of oxygen, respectively, where the dominant contribution arises from the interfacial Pt atoms. We find that in general $\langle\hat{\xi}{\hat{L}_y}\rangle_{0,\sigma}>$0 and  that
 $\langle\hat{\xi}\vec{\hat{L}}\rangle_{0,\downarrow}$ is insensitive to the presence or absence of oxygen. Consequently, 
the sign reversal of the FL-SOT is due to the significant reduction of the majority-spin contribution induced by interfacial oxygen.

Furthermore, Figs.~\ref{fig:fig4}(c) and (f) show the orbital-resolved contributions to $B_{y}^{REE}$ for the interfacial Pt atom, without and with oxygen, respectively. The dominant contribution to the 
$\langle{Ilm\sigma}|B_{y}^{REE}|{Ilm'\sigma'}\rangle\propto\sigma\xi_{Il}
\langle{lm}|\hat{L}_{y}|lm'\rangle$,
arises from the non-vanishing $\langle d_{xy}|\hat{L}_{y}|d_{yz}\rangle$, $\langle d_{xz}|\hat{L}_{y}|d_{z^2}\rangle$ and  $\langle d_{xz}|\hat{L}_{y}|d_{x^2-y^2}\rangle$  matrix elements of the in-plane orbital angular momentum operator, $\hat{L}_{y}$. Here, $I,\sigma,l,m$ stand for ionic, spin and atomic orbital indices, respectively.
We find that the
positive  $\langle d_{xy}|\hat{L}_{y}|d_{yz}\rangle$ and $\langle d_{xz}|\hat{L}_{y}|d_{z^2}\rangle$ matrix elements in the clean bilayer vanish or change sign by oxygen, which can be attributed to the hybridization of 
the majority-spin O/$p$-derived states with the interfacial  Pt/$\{d_{xz}$,$d_{yz}$ and $d_{z^2}\}$  derived states.\cite{Supp_Matt}


In summary, we presented a theoretical framework which allows to decompose the DL- and FL-SOTs into
the spin-Hall and spin-orbital components in a NM/FM bilayers.  We demonstrated that (i) the FL-SOT is dominated by the spin-orbital component which can be expressed in terms of the difference of nonequilibrium spin-resolved orbital accumulation at the interfacial NM atoms and (ii) the spin-orbital contribution to the DL-SOT dominates for ultra-thin Pt films, while both components have equal contribution for thicker films. The dependence of the DL-SOT on SOC strength suggests that if Pt is replaced with a normal metal with weaker SOC strength, the DL-SOT becomes dominated by the spin-Hall component.
We have used the approach to elucidate the microscopic mechanism for the sign reversal of the FL-SOT due to the interfacial Co oxidation. We demonstrated that the sign reversal is attributed to a significant reduction of the spin-majority nonequlibrium orbital moment accumulation at the interfacial Pt layer.

	
	
\end{document}